\documentclass[journal=jpclcd,manuscript=letter]{achemso}

\usepackage{graphicx}

 \usepackage{subfigure}
 \usepackage{csquotes}
 \usepackage{amssymb}

\author{Leonardo del Rosso}
\affiliation{Consiglio Nazionale delle Ricerche, Istituto dei Sistemi Complessi, via Madonna del Piano 10, I-50019 Sesto Fiorentino, Italy}
\alsoaffiliation{Dipartimento di Fisica e Astronomia, Universit\`a degli Studi di Firenze, via Sansone 1, I-50019 Sesto Fiorentino, Italy}

\author{Francesco Grazzi}
\affiliation{Consiglio Nazionale delle Ricerche, Istituto dei Sistemi Complessi, via Madonna del Piano 10, I-50019 Sesto Fiorentino, Italy}

\author{Milva Celli}
\affiliation{Consiglio Nazionale delle Ricerche, Istituto dei Sistemi Complessi, via Madonna del Piano 10, I-50019 Sesto Fiorentino, Italy}

\author{Daniele Colognesi}
\affiliation{Consiglio Nazionale delle Ricerche, Istituto dei Sistemi Complessi, via Madonna del Piano 10, I-50019 Sesto Fiorentino, Italy}

\author{Victoria Garcia-Sakai}
\affiliation{ISIS Pulsed Neutron and Muon Source, STFC Rutherford Appleton Laboratory, Chilton, Oxfordshire OX11 0QX, U.K.}

\author{Lorenzo Ulivi}
\email{lorenzo.ulivi@isc.cnr.it} 
\affiliation{Consiglio Nazionale delle Ricerche, Istituto dei Sistemi Complessi, via Madonna del Piano 10, I-50019 Sesto Fiorentino, Italy}

\title{Refined Structure of Metastable Ice XVII from Neutron Diffraction Measurements}


\begin{document}

 \begin{tocentry}
 \resizebox{5cm}{!}
 {
  \includegraphics[bb=0cm 0cm 27.9cm 21cm, viewport= 4.5cm 0.1cm 24cm 21cm]
   {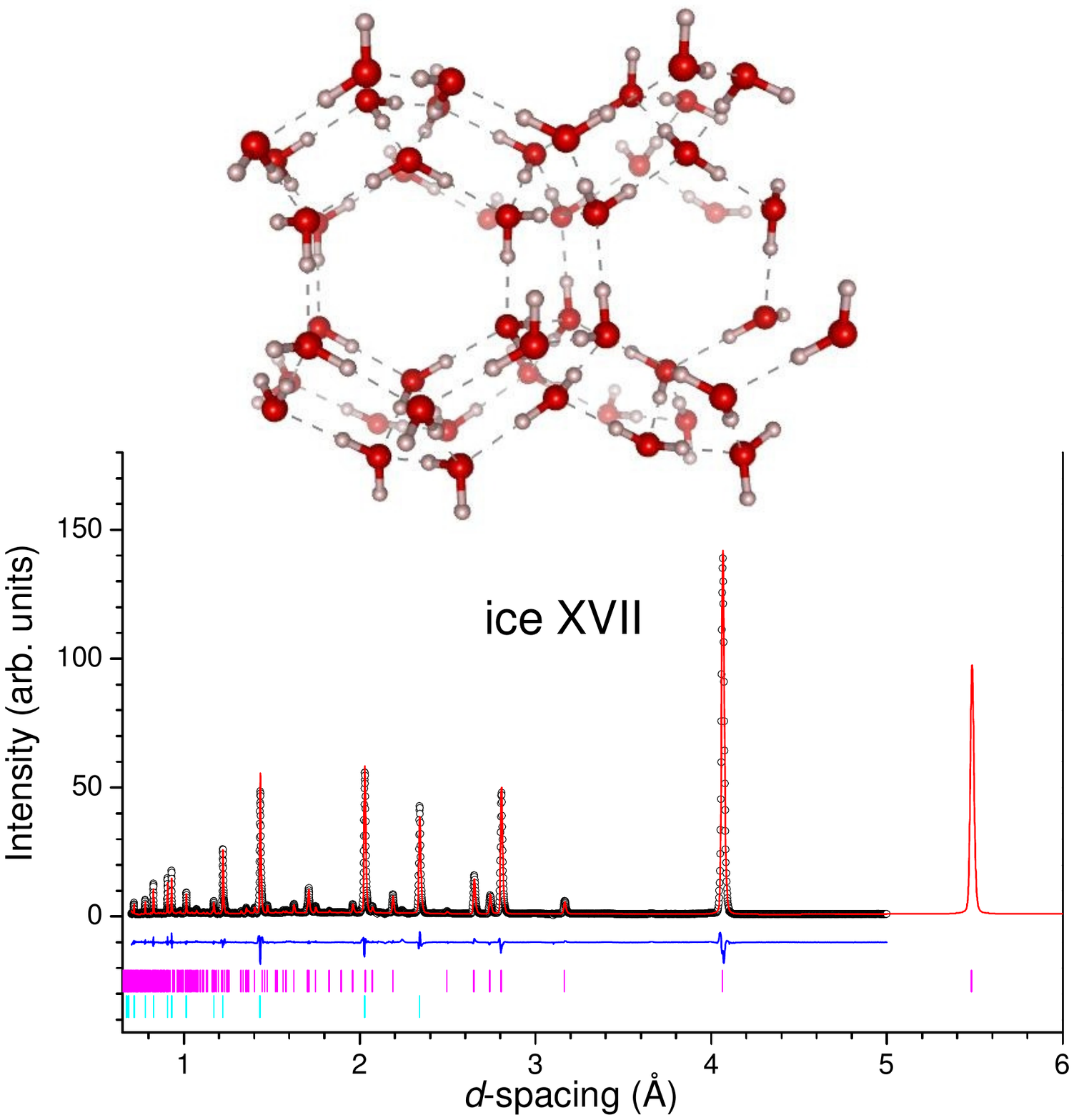}
 }

 \end{tocentry}

\begin{abstract} 
 
The structure of the recently identified metastable ice XVII, obtained by release of hydrogen from the C$_0$ D$_2$O-H$_2$ compound (filled ice), has been accurately measured by neutron powder diffraction.
The diffraction pattern is indexed with a hexagonal cell and can be refined with space group $P6_122$ so to obtain accurate values of the oxygen and deuterium positions.
The values of the lattice constants at three temperatures between 25 to 100 K are reported, and their behavior is compared with that of ice Ih. 
Ice XVII is a porous solid that, if exposed to H$_2$ gas,  may adsorb a substantial amount of it. 
Monitoring this effect at a constant temperature of 50 K, we have observed that  the two lattice constants show opposite behavior,  $a$ increases and $c$ decreases, with the volume  showing a linear increase.
At temperatures higher than 130 K the metastability of this form of porous ice is lost and the sample transforms into ice Ih.   
 
\begin{center}
\date{\today}
\end{center}

\end{abstract}

Solid phases formed by mixtures of water and hydrogen give rise to various forms of crystal structures, where the guest H$_2$ molecules are trapped either inside cages of different geometries, as in clathrate hydrates\cite{Sloan08,Dyadin99,MaoW02}, or intercalated in the structure, as in other high pressure compounds known as  \enquote{filled ices} \cite{Vos93,Efimchenko11,Strobel11}.
In the clathrate-hydrate structure, which is cubic of type sII, the hydrogen molecules inside the cages exhibit a quantum dynamics which has been investigated efficiently by inelastic neutron scattering\cite{Ulivi07,Xu11a,Xu13a,Celli13,Colognesi13} and Raman scattering\cite{Giannasi08a,Strobel09,Giannasi11,Zaghloul12}.
Only recently it has been demonstrated that, contrary to the belief that caged guest molecules are essential for the stability of the clathrate structure, some instance of these may exist in a metastable form at low temperature even in the absence of guest molecules.\cite{Falenty14}   
In a recent paper we have shown that a similar phenomenon happens also for the C$_0$ filled-ice.\cite{delRosso16}
This crystal is the stable equilibrium phase of the mixture water-hydrogen at temperatures 100-270 K and pressures 360-700 MPa, that is intermediate between the stability region of sII clathrate and that of C$_1$ filled ice\cite{Efimchenko11,Strobel11}. 
Once synthesized at high pressure from water and hydrogen, C$_0$ filled ice can be recovered at room pressure and low temperature, still containing a large fraction of molecular hydrogen.
The recovered sample releases all its hydrogen by heating under vacuum for about one hour at a temperature of 110-120 K, giving rise to a new form of metastable solid phase of water, which we have named ice XVII.\cite{delRosso16}
This exotic and low-density solid structure of water molecules adds to the  list of solid structures of water possibly stable at negative pressure.\cite{Huang16}.
In this paper we present the results of a neutron diffraction measurement performed on OSIRIS at ISIS, RAL (UK)  whose analysis enabled us to determine and refine the structure of ice XVII.
 
We have produced the sample in the C$_0$ phase by using heavy water and normal hydrogen.
Ground D$_2$O ice has been inserted into a beryllium-copper autoclave, and a pressure above 430 MPa of hydrogen has been applied at a temperature of about 255 K (-18$^{\circ}$C).
The sample has been maintained under pressure for a few days,
after which it has been quenched at liquid nitrogen temperature and recovered in the form of a fine powder.
The check of the recovered sample by Raman spectroscopy has confirmed that we have obtained the C$_0$ phase, excluding the presence of sII clathrate or other ice phases.\cite{delRosso16}
We have used an aluminum gas cell having the shape of a hollow cylinder and provided with a valve.
The transfer process of the sample into the cell has been accomplished at liquid nitrogen temperature in a dry-nitrogen atmosphere, exerting particular care to avoid sample heating and contamination from air water vapor.
The cell is transfered to ISIS, RAL (UK),  inserted in the neutron cryostat and connected through a capillary to a vacuum and gas control system.
The OSIRIS time-of-flight instrument allows simultaneous Quasi-Elastic-Neutron-Scattering (QENS) and diffraction measurements.
For diffraction, OSIRIS uses a backscattering detector bank consisting of 962 ZnS scintillators with an angle coverage $150\,^{\circ} < 2\, \theta <171\,^{\circ}$.
The covered $d$-range spans between 0.8 \AA\ and $\approx 20$ \AA\, with a high resolution, namely $2.5 \cdot10^{-3} < \frac{\Delta d}{d} < 6.0 \cdot 10^{-3}$.
Counting rate is high and allows a diffraction measurement in the range of interest (0.8 \AA\ $< d < 6.0$ \AA)  in about 30-60 minutes.
The experiment was aimed to measure also diffusion of guest hydrogen by means of  QENS.
%
That is why H$_2$ instead of D$_2$ is used as guest gas.
Diffraction data  have been collected in the whole  range of $d$ only in some instances, while diffraction patterns in the limited range 3.0 \AA\ $ <d< 4.3 $ \AA\ were recorded more often for a fast check of the sample.
The use of hydrogen as guest gas hampers the refinement of the positions of the guest molecules in the C$_0$ crystal, but data from the empty ice XVII will serve perfectly for the refinement of its structure.

\begin{figure}[!h]
\begin{center}
\resizebox{0.8\textwidth}{!}
{
\includegraphics[bb=0cm 0cm 27.9cm 21cm, viewport= 1cm 1cm 27cm 22cm]
{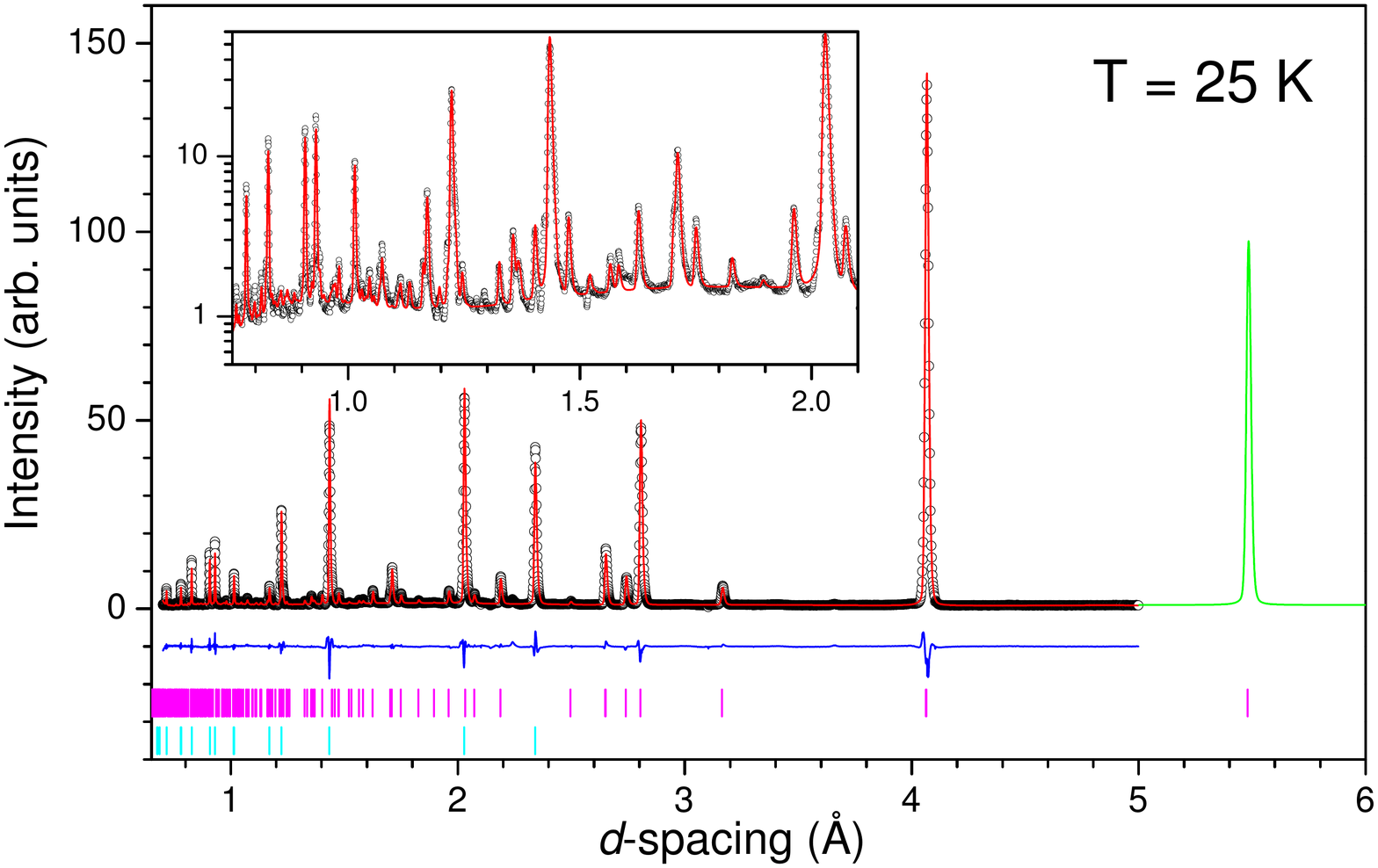}
}
\end{center}
\caption{(Color online) Diffraction pattern recorded at 25 K used for the refinement of the structure with the Rietveld method.
Space group used is $P6_122$.  
The refined parameters are listed in tab.~\ref{t.parameters}.
In the inset an enlargement of the low $d$ zone is presented on a semi-log plot. 
Circles are experimental points, fitted by the red line, while residuals are in blue.
Magenta and cyan vertical bars mark the positions of ice XVII and aluminum peaks, respectively.
For $d> 5 $ \AA\  experimental data are not fitted, and the calculated diffraction peak is presented in the figure (green line).
}
\label{f.diffraction}
\end{figure}

\begin{figure}[!h]
\begin{center}
\resizebox{0.8\textwidth}{!}
{
\includegraphics[viewport= 0.5cm 0.5cm 21.5cm 16.5cm]
{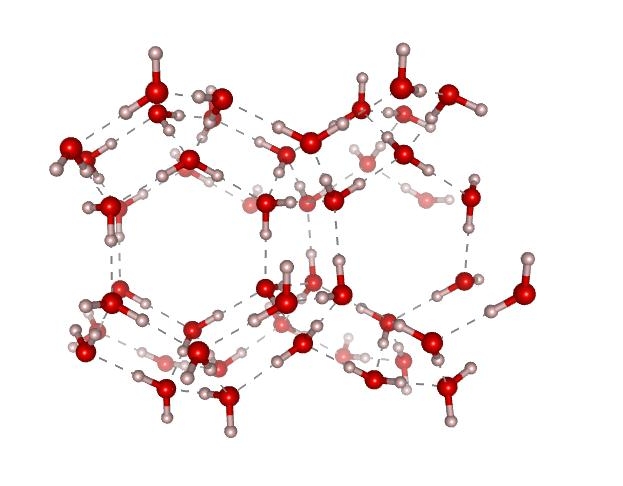}
}
\end{center}
\caption{(Color online) Prospectic  drawing of the structure of ice XVII  viewed along the $z$ axis. The channels have a free bore hole along the $z$ axis of about 5.26 \AA\ and with a diameter of 6.10 \AA\ (produced with software VESTA\cite{Vesta11}).
}
\label{f.structure}
\end{figure}

After a few measurements on the pristine C$_0$ sample at increasing temperature, the cell is heated to 100 K under vacuum for a few hours, in order to obtain ice XVII.
This annealing process assures the removal of the guest H$_2$ molecules and nitrogen molecules that are often present in the sample prepared with the procedure described above\cite{delRosso16}.
Full range diffraction patterns of ice XVII are then collected at three temperatures, namely 100, 75,  and 25 K, before the sample is let adsorb hydrogen again.
%
Data at all the three temperatures have been analyzed by Rietveld refinement using the software GSAS\cite{Larson00}.
Preliminarily, in order to obtain a reliable profile fit function and other instrumental parameters, we have performed a Rietveld refinement on an NaCaAlF diffraction pattern, measured a shortly before our experiment and available in the instrument data base, obtaining the best fit with a time-of-flight profile function of type 4\cite{Larson00}.
For the Rietveld refinement of our data we have excluded the range $d > 5$ \AA\, because of the unreliability of the peak intensity and peak shape model in this long wavelength region.
As a matter of fact, the beam intensity behavior for $\lambda > 10$ \AA\ is peculiar to the hydrogen moderator since a large number of neutron collisions inside it are necessary.
This phenomenon strongly affects the pulse shape and none of the peak profile functions available in GSAS is able to model it correctly.
Moreover, also the peak profile shape extracted by mean of the NaCaAlF refinement is somewhat inadequate for large $\lambda$.
This sample is a common one as a calibration standard for its wide $d$-range span, but it is not ideal because of its variable synthesis conditions, bringing to the presence of several micro-structural features affecting the diffraction peak shape, such that defect density, domain size, and compositional discrepancies.
Anyhow, in the excluded range only one peak ($hkl = 100$) of the diffraction pattern of ice XVII is present and in the whole remaining region the pulse shape can be modeled with adequate accuracy so that  peak profiles in NaCaAlF are well reproduced by profile of type 4.
The space groups assumed for the Rietveld refinement of ice XVII diffraction pattern are those proposed for describing the filled structure C$_0$, namely  $P3_112$\cite{Efimchenko11} and $P6_122$\cite{Strobel16}.
The quality of the fit, estimated from the value of the parameter $R_{wp}$, turns out to be similar ($R_{wp} \simeq 11 \%$ in both cases), but the values obtained for the oxygen coordinates in sites $3a$ and $3b$ of the group $P3_112$ are related by $x_{3a} \approx 1-x_{3b}$, so that these positions coincide with that of the site $6b$ of the group $P6_122$.
In addition, having the group $P6_122$ a higher symmetry and a lower number of free parameters, it is to be preferred for the description of the structure.
The experimental pattern and the refined model at 25 K is reported in fig.~\ref{f.diffraction}, while their refined structural parameters are reported in tab.~\ref{t.parameters}.
The refined positions of D atoms obtained from the fit are perfectly consistent with the water molecule structure, which is reproduced without the need of any constraint.
Nearest neighbor oxygens are tetrahedrally coordinated and sit at two possible distances, namely $d_1=2.7405$ and  $d_2=2.7733$.
Corresponding to these, O-D distances are 1.020 and 1.002 \AA, respectively.
Four possible values are obtained for the O-O-O and D-O-D angles, listed in tab.~\ref{t.angles}, which are correlated with the distance, $d_1$ or $d_2$, of each O-O pair.
There is a large correlation between O-O-O and D-O-D angles, but the water molecule tends to possess an angle whose average is 110.0 degrees, while O-O-O angle is in some instances as large as 124.5 degrees.
A picture of the structure of ice XVII is shown in fig.~\ref{f.structure}. 
Water molecules form spiraling channels with a free bore hole along the $z$ axis of about 5.26 \AA\ and with a diameter of 6.10 \AA ,  which can accommodate H$_2$ molecules.

\begin{figure}[!h]
\begin{center}
\resizebox{0.8\textwidth}{!}
{
\includegraphics[viewport= 2cm 2cm 20cm 27cm]
{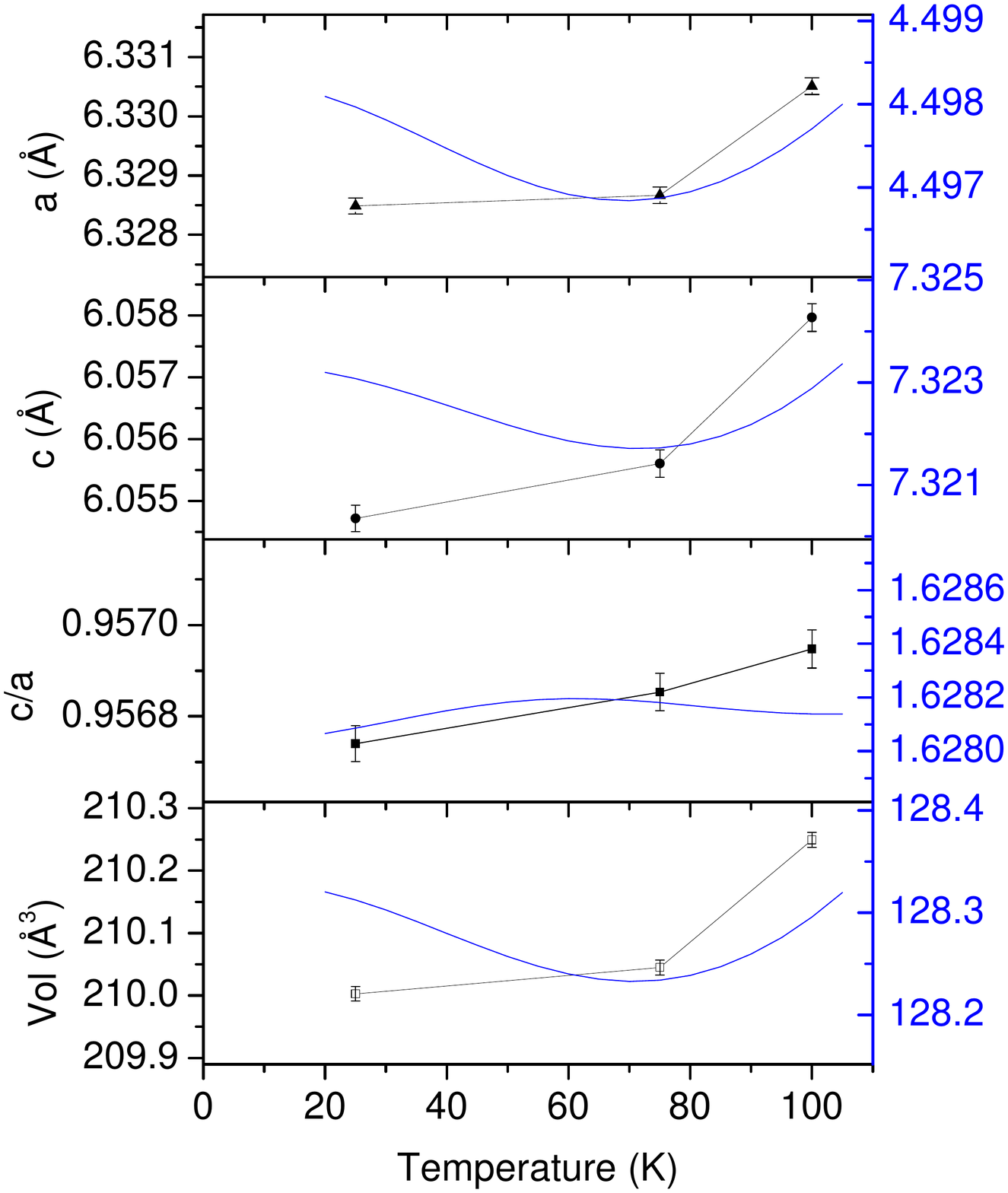}
}
\end{center}
\caption{(Color online) Temperature dependence of the lattice constants of D$_2$O ice XVII (black symbols and black line, left vertical scale) and of D$_2$O ice Ih\cite{Rottger94} (blue line, right vertical scale).
To help the comparison, the relative vertical scale range of the right and left axis is the same, and correspond to the same relative change of the parameter, (namely 3.5, 3.5, 3.0, 10.0 $10^{-4}$ for $a$, $c$, $c/a$ and Vol respectively).  
}
\label{f.vsT}
\end{figure}

Data collected at the other two temperatures are refined with similar fit quality, and show that the structure does not change with temperature.
The behavior with temperature of  lattice constants and unit cell volume is reported in fig.~\ref{f.vsT}, where it is compared with that of ice Ih\cite{Rottger94}.
The density of the crystal at 100 K is  0.95018(5) g/cm$^3$, that is 11\% lower than that of D$_2$O ice Ih at the same temperature. 
The low number of investigated temperatures does not allow us to determine  with precision the thermal expansivity,  which may be possibly negative between 25 and 75 K.
Anyhow, thermal expansion is not isotropic in ice XVII,  in contrast to ice Ih, as it is testified by observing the $c/a$ ratio in fig.~\ref{f.vsT}.

\begin{figure}[!h]
\begin{center}
\resizebox{.9\textwidth}{!}
{
\includegraphics[viewport= 1cm 1cm 27cm 20cm]
{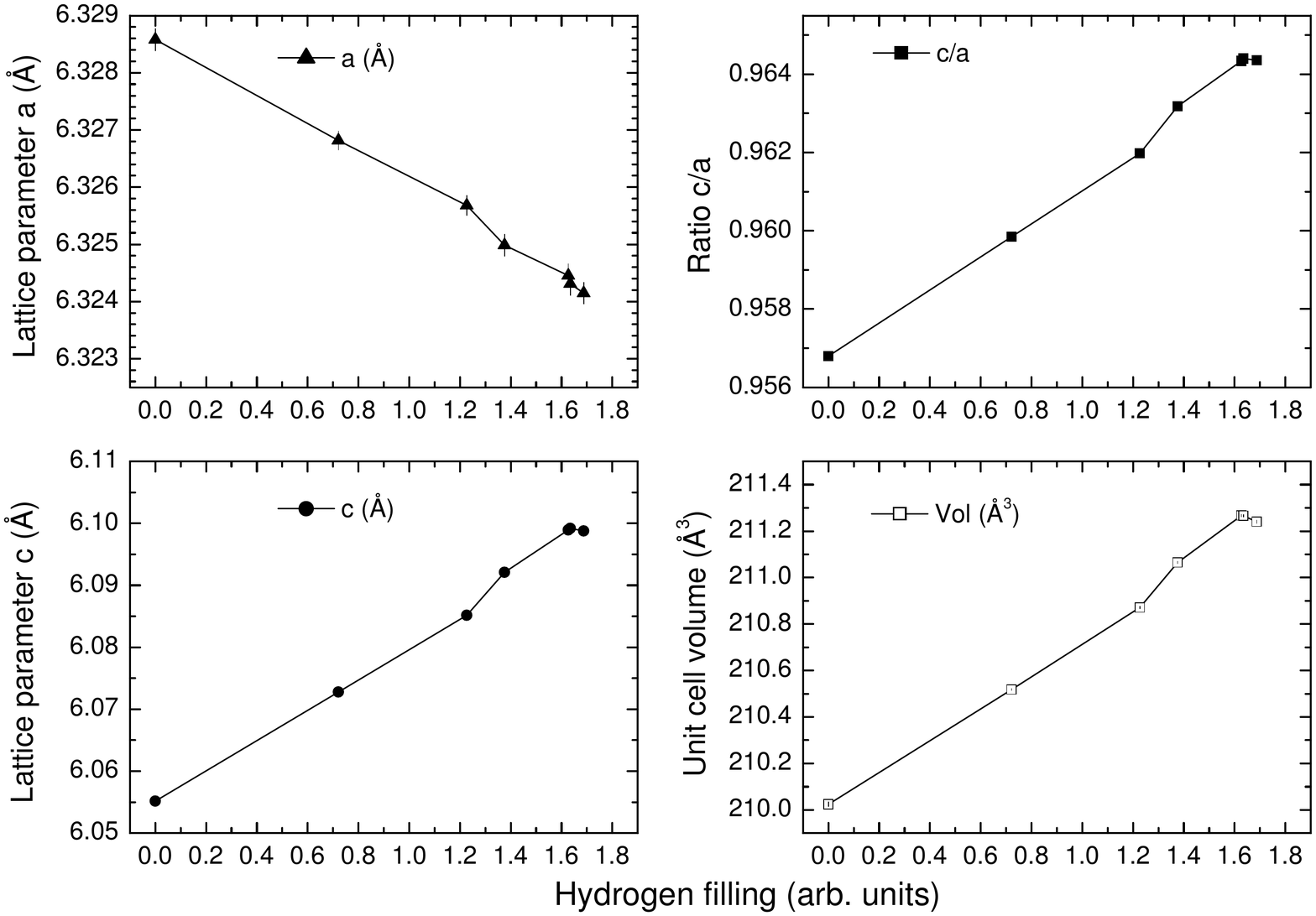}
}
\end{center}
\caption{ Dependence of the lattice constants of the C$_0$ filled ice upon filling with hydrogen, at 50 K.
The amount of hydrogen adsorbed in the sample is given in arbitrary units and is estimated from the incoherent background present in the diffraction data, which in these cases span a limited range in $d$, namely between 3 and 4.3 \AA. 
}
\label{f.vsH2}
\end{figure}

After the measurement of the empty sample we have checked its ability to re-adsorb H$_2$, and measured diffraction patterns in a limited $d$ range.
These data, even if not  extensive enough to be suitable for Rietveld refinement, give us access to the lattice constant dependence as a function of the H$_2$ content.
This quantity has not been measured independently, but has been estimated from diffraction data.
It is well known, in fact, that the large incoherent scattering cross section of H$_2$ molecule gives rise, in the diffraction pattern, to a significant background, almost independent of $d$, which is easily measurable and can be considered proportional to the adsorbed H$_2$.
Even if we cannot give an estimate of the absolute hydrogen content, this relative calibration allows us to plot lattice parameter behavior in function of hydrogen content and discuss the graphs shown in fig.~\ref{f.vsH2}.
Upon filling, we notice a 0.6 \% volume increase, which however cannot  be naively attributed to a swelling of the ice due to the guest, since H$_2$-D$_2$O interaction is mainly attractive for the range of distance pertinent to the hosting of  H$_2$ in the channels, as it is attractive for Ne in the sII clathrate\cite{Falenty14}.
As a matter of fact, the lattice constant $a$ decreases with adsorption, and the increase of volume is due to the elongation of the unit cell along the $z$ axis.
We want to underline the fact that these changes are 6-10 times larger than those due temperature change in the range 25-100 K.

\begin{figure}[!h]
\begin{center}
\resizebox{.9\textwidth}{!}
{
\includegraphics[viewport= 1cm 1cm 27cm 20cm]
{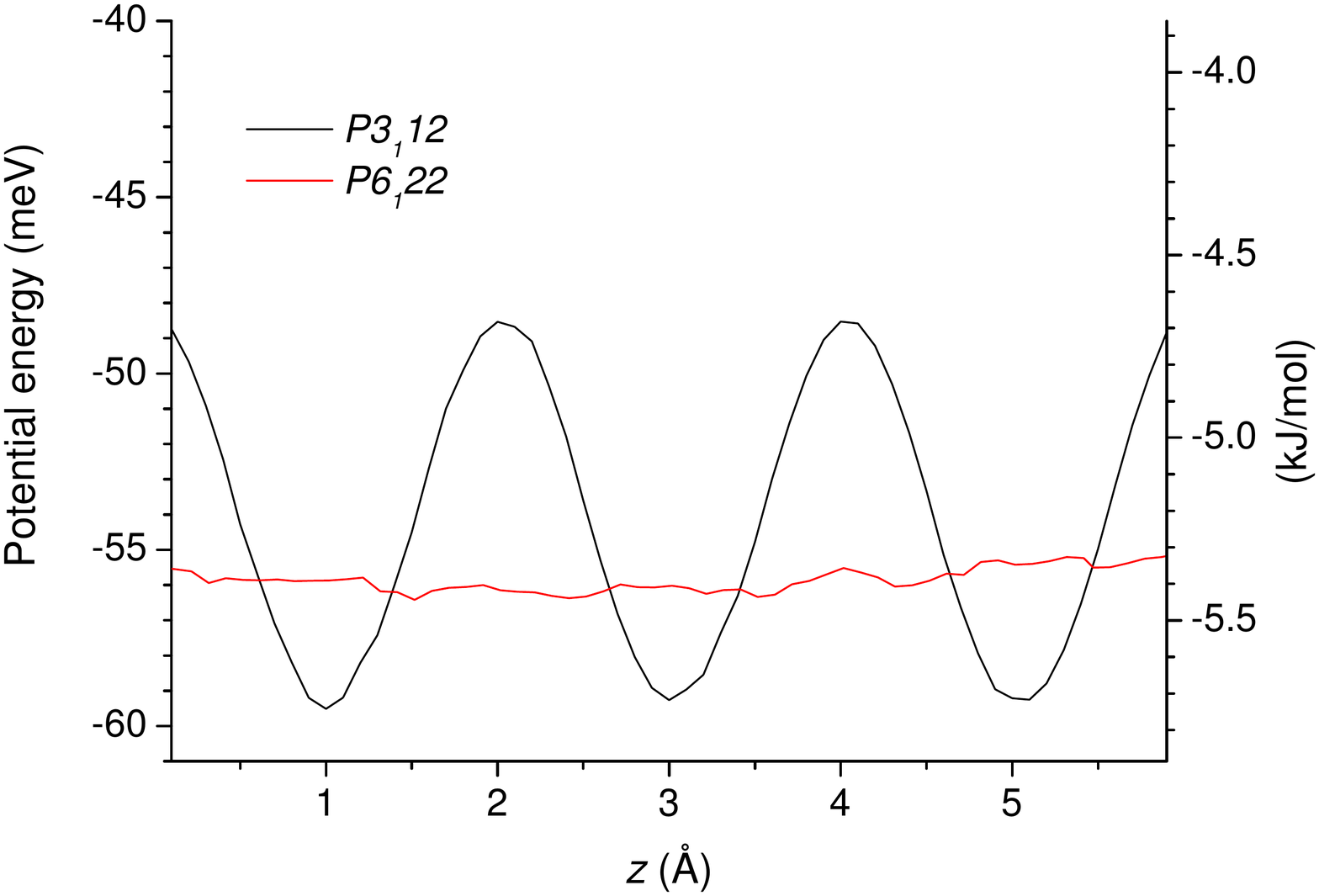}
}
\end{center}
\caption{Potential energy barriers for the motion of one H$_2$ molecule along the helix of relative minima inside the channels, for the two  structures considered.
The right vertical scale reports the potential energy in kJ/mol.
Lattice constants assumed for this calculation are $a=6.331396$ and $c=  6.055204$, and site coordinates $x_{3a}=0.23$ $x_{3b}=0.75$ for group $P3_112$  and $x_{6b}=0.24$  for group $P6_122$.
Potential energy model for H$_2$-D$_2$O interaction is based on the SPC/e model for water, interacting with a Lennard-Jones site on the H$_2$ center of mass and on point charges on D, H, O and center of mass of H$_2$, as in ref.~\citenum{Alavi05}.   
}
\label{f.helix}
\end{figure}

The data we have collected do not comprise the diffraction patterns of the refilled sample in the complete range of $d$, so we cannot draw any conclusion on the water lattice structure  of the C$_0$ based on precise diffraction data. 
We can however discuss the difference of the two proposed structures, $P6_122$ and  $P3_112$, and  speculate on what could be the effect of the guest from a theoretical point of view.
Space group $P3_112$   is a subgroup of $P6_122$. 
One structure can change into the other by means of a displacive transition. For particular values of the coordinates of the atoms in sites $3a$ and $3b$ the symmetry of the structure increases and the crystal can be described by the group $P6_122$.
We have calculated the potential energy field probed by an H$_2$ molecule as a function of its position inside the channel, for the two structures considered as rigid.
For each structure, we have located in the $xy$ plane the minimum of the energy  for about 60 fixed values of the vertical coordinate $z$.
These minima are located along a helix, winding around a cylinder with a radius of about 2.0 \AA, having the same period as the lattice along $z$.
The potential energy along the helices, however, has a  quite different behavior for the two structures.
While for $P6_122$ a hypothetical H$_2$ molecule moving along the   helix would feel an almost constant potential energy, for $P3_112$ the same molecule should cross three potential maxima (and three minima) per unit cell length.
The situation is depicted in fig~\ref{f.helix}. 
The potential barriers between the minima ($\simeq 10$ meV) are not so high to prevent diffusion of H$_2$, since zero-point energy is probably higher.
Given the low energy difference between the two structure, it is also possible that H$_2$ guests, entering the channels, modify the symmetry of the structure of ice XVII, and accommodate themselves at the bottom of the potential well that themselves induce.
In this picture the H$_2$ molecules would occupy sites with a spacing commensurate with the period of the host lattice.
Two observation are in agreement with this speculation.
One is the large hysteresis in the gas absorption-release process, described in detail in our previous paper\cite{delRosso16}, and the maximum amount of adsorbed H$_2$ that gets close to 50\% molar.
The other is the elongation of the unit cell along the $c$, axis, that might be  favored by the repulsive potential energy between nearest-neighbor guests,     
which would influence the structure.
More measurements are needed to confirm this speculation.

 

%

\vspace{5cm}

\begin{table}[h!]
\caption{Atomic fractional coordinates obtained by means of Rietveld refinement (space group $P6_122$) of the diffraction pattern of ice XVII measured at 25 K. Lattice constants are $a=6.32849(14)$ \AA, $c=6.05472(22)$ \AA} 
\centering 
\begin{tabular}{|c| c| c| c|c|c| c|} 
\hline\hline 
 Atom & site  & $x$ &  $y$  & $z$  & B$_{iso}$   & Occ.       \\  
\hline 
O1 & $6b$ & 0.23578(37)   &   $2x$  &   $ 1/4 $            &      2.37(10)   &        1.0 \\
D1 & $12c$  &  0.6573(10)  &  0.42197(95)    &   0.8769(13) &    4.05(16)    &       0.5  \\
D2 &  $12c$ &  0.93633(99)   &   0.62954(72)   &  0.80539(92)    &   3.34(14)     &      0.5  \\ 
 \hline 

\end{tabular}
\label{t.parameters} 
\end{table}
\begin{table}[h!]
\caption{Atomic distances and angles obtained by the Rietveld refinement (space group $P6_122$) of the diffraction pattern of ice XVII measured at 25 K.} 
\centering 
\begin{tabular}{|c| c| c| } 
\hline\hline 
 O-O pairs distance   & O-O-O & D-O-D     \\  
    & degrees  & degrees   \\
\hline 
$d_1$-$d_1$  & 108.28  & 110.21   \\
$d_2$-$d_2$  &  124.49 &  113.28  \\
$d_1$-$d_2$  &  106.12 &  108.60  \\
$d_2$-$d_1$  &  105.54 &  108.07  \\
 \hline 

\end{tabular}
\label{t.angles} 
\end{table}

\begin{acknowledgement}
The STFC Rutherford Appleton Laboratory is thanked for
access to neutron beam facilities.
Fruitful discussion with Prof.~Michele Catti, (University of Milan, Italy) are gratefully acknowledged.
\end{acknowledgement}

\begin{suppinfo}

Crystallographic information file (CIF) of the refined structure of ice XVII at 25, 75 and 100 K.
\end{suppinfo}
\newpage

\newpage
%
\newpage
\newpage
%
\newpage
%
%
\newpage


\begin{thebibliography}{00}

\bibitem{Sloan08} Sloan, E. D.; Koh, C. A.; {\it Clathrate Hydrates of Natural Gases, 3$^\mathrm{rd}$ Edition}; Taylor \& Francis, New York, 2008.
%
\bibitem{Dyadin99} Dyadin, Y. A.; Larionov, E. G.; Manakov, A. Y.; Zhurko, F. V.; Aladko, E. Y.; Mikina, T. V.; Komarov, V. Y. Clathrate Hydrates of Hydrogen and Neon. {\it Mendeleev Commun.} {\bf 1999}, {\it 9}, 209-210.
%
\bibitem{MaoW02} Mao, W. L.; Goncharov, A. F.; Struzhkin, V. V.; Guo, Q.; Hu, J.; Shu, J.; Hemley, R. J.; Somayazulu, M.; Zhao, Y. Hydrogen Clusters in Clathrate Hydrate. {\it Science} {\bf 2002}, {\it 297}, 2247-2249.
%
\bibitem{Vos93} Vos, W. L.; Finger, L. W.; Hemley, R. J.; Mao, H. K. Novel H$_2$+H$_2$O Clathrates at High Pressures. {\it Phys. Rev. Lett.} {\bf 1993}, {\it 71}, 3150-3153.
%
\bibitem{Efimchenko11} Efimchenko, V. S.; Kuzovnikov, V. S.; Fedotov, V. K.; Sakharov, M. K.; Simonov, S. V.; Tkacz, M. New Phase in the Water-Hydrogen System. {\it J. Alloys Comp.} {\bf 2011}, {\it 509}, S860-S863.
%
\bibitem{Strobel11} Strobel, T. A.; Somayazulu, M.; Hemley, R. J. Phase Behavior of H$_2$+H$_2$O at High Pressures and Low Temperatures. {\it J. Phys. Chem. C} {\bf 2011}, {\it 115}, 4898-4903. 
%
\bibitem{Ulivi07} Ulivi, L.; Celli, M.; Giannasi, A.; Ramirez-Cuesta, A. J.; Bull, D. J.; Zoppi, M. Quantum Rattling of Molecular Hydrogen in Clathrate Hydrate Nanocavities. {\it Phys. Rev. B} {\bf 2007}, {\it 76}, 1614017/1-1614017/4. 
%
\bibitem{Xu11a} Xu, M.; Ulivi, L.; Celli, M.; Colognesi, D.; Ba\v ci\' c, Z. Quantum Calculation of Inelastic Neutron Scattering Spectra of a Hydrogen Molecule Inside a Nanoscale Cavity Based on Rigorous Treatment of the Coupled Translation-Rotation Dynamics. {\it Phys. Rev. B} {\bf 2011}, {\it 83}, 241403/1-241403/4. 
%
\bibitem{Xu13a} Xu, M.; Ulivi, L.; Celli, M.; Colognesi, D.; Ba\v{c}i\'{c}, Z. Rigorous Quantum Treatment of Inelastic Neutron Scattering Spectra of a Heteronuclear Diatomic Molecule in a Nanocavity: HD in the Small Cage of Structure II Clathrate Hydrate. {\it Chem. Phys. Lett.} {\bf 2013}, {\it 563}, 1-8.
%
\bibitem{Celli13} Celli, M.; Powers, A.; Colognesi, D.; Xu, M.; Ba\v{c}i\'{c}, Z.; Ulivi, L. Experimental Inelastic Neutron Scattering Spectrum of Hydrogen Hexagonal Clathrate-Hydrate Compared with Rigorous Quantum Simulations. {\it J. Chem. Phys.} {\bf 2013}, {\it 139}, 164507/1-164507/8. 
%
\bibitem{Colognesi13} Colognesi, D.; Celli, M.; Ulivi, L.; Xu, M.; Ba\v{c}i\'{c}, Z. Neutron Scattering Measurements and Computation of the Quantum Dynamics of Hydrogen Molecules Trapped in the Small and Large Cages of Clathrate Hydrates. {\it J. Phys. Chem. A} {\bf 2013}, {\it 117}, 7314-7326.
%
\bibitem{Giannasi08a} Giannasi, A.; Celli, M.; Ulivi, L.; Zoppi, M. Low Temperature Raman Spectra of Hydrogen in Simple and Binary Clathrate Hydrates. {\it J. Chem. Phys.} {\bf 2008}, {\it 129}, 084705/1-084705/10.
%
\bibitem{Strobel09} Strobel, T. A.; Sloan, E. D.; Koh, C. A. Raman spectroscopic studies of hydrogen clathrate hydrates {\it J. Chem. Phys.} {\bf 2009}, {\it 130}, 014506/1-014506/10. 
%
\bibitem{Giannasi11} Giannasi, A.; Celli, M.; Zoppi, M.; Moraldi, M.; Ulivi, L. Experimental and Theoretical Analysis of the Rotational Raman Spectrum of Hydrogen Molecules in Clathrate Hydrates. {\it J. Chem. Phys.} {\bf 2011}, {\it 135}, 054506/1-054506/9. 
%
\bibitem{Zaghloul12} Zaghloul, M. A. S.; Celli, M.; Salem, N. M.; Elsheikh, S. M.; Ulivi, L. High Pressure Synthesis and In Situ Raman Spectroscopy of H$_2$ and HD Clathrate Hydrates. {\it J. Chem. Phys.} {\bf 2012}, {\it 137}, 164320/1-164320/8.
%
\bibitem{Falenty14} Falenty, A.; Hansen, T. C.; Kuhs, W. F. Formation and Properties of Ice XVI Obtained by Emptying a Type sII Clathrate Hydrate. {\it Nature} {\bf 2014}, {\it 516}, 231-233.
%
\bibitem{delRosso16} del Rosso, L.; Celli, M.; Ulivi, L. A New Porous Water Ice Stable at Atmospheric Pressure Obtained by Emptying a Hydrogen Filled Ice. {\it Nature Comm. (submitted)} {\bf 2016}, arXiv:1607.07617  https://arxiv.org/abs/1607.07617.
%
\bibitem{Huang16} Huang, Y.; Zhu, C.; Wang, L.; Cao, X.; Su, Y.; Jiang, X.; Meng, S.; Zhao, J.; Zeng, X. C.; A New Phase Diagram of Water Under Negative Pressure: The Rise of the Lowest-Density Clathrate s-III. {\it Science Advances} {\bf 2016}, {\it 2}, e1501010/1-e1501010/6. DOI: 10.1126/sciadv.1501010 
%
\bibitem{Larson00} Larson, A. C.; Von Dreele, R. B. General Structure Analysis System (GSAS). {\it Los Alamos National Laboratory Report LAUR} {\bf 2000}, 86-748.
%
\bibitem{Strobel16} Strobel, T. A.; Somayazulu, M. S.; Sinogeikin, S. V.; Dera, P.; Hemley, R. J. Hydrogen-Stuffed, Quartz-Like Water Ice. {\it J. Am. Chem. Soc.}, {\bf 2016} (Just Accepted Manuscript). DOI: 10.1021/jacs.6b06986
%
\bibitem{Rottger94} Rottger, K.; Endriss, A.; Ihringer, J.; Doyle, S.; Kuhs, W. F. Lattice Constants and Thermal Expansion of H$_2$O and D$_2$O Ice Ih Between 10 and 265 K. {\it Acta Cryst.} {\bf 1994}, {\it B50}, 644-648 and addendum {\it Acta Cryst.} {\bf 2012}, {\it B68}. 
%
\bibitem{Vesta11} Momma, K.; Izumi, F. VESTA 3 for Three-Dimensional Visualization of Crystal, Volumetric and Morphology Data. {\it J. Appl. Crystallogr.} {\bf 2011}, {\it 44}, 1272-1276.
%
\bibitem{Alavi05} Alavi, S.; Ripmeester, J. A.; Klug, D. D. Molecular-Dynamics Study of Structure II Hydrogen Clathrates. {\it J. Chem. Phys.} {\bf 2005}, {\it 123}, 024507/1-024507/7.








%
%
%
%
%
%
%
%
%
%
%
%
%
%
%
%
%
%
%
%
%
%
%
%
%
%
%
%
%
%
%
%
%
%
%
%
%
%
%
%
%
%
%
%
%
%

%
%
%
%




\end{thebibliography}
\end{document}